\documentclass[twocolumn,aps,prb,10pt,nofootinbib,superscriptaddress]{revtex4-2}

\usepackage{amsmath}
\usepackage{amssymb}
\usepackage{wasysym}
\usepackage{graphicx}
\usepackage{hyperref}
\usepackage{dsfont}
\usepackage{mathtools}
\usepackage{newtxtext}
\usepackage[varvw]{newtxmath}

\hypersetup{
    colorlinks,
    linkcolor={blue},
    citecolor={blue},
    urlcolor={blue}
}

\graphicspath{{./}{./figs/}}

\newcommand{\rme}{\mathrm{e}}
\newcommand{\rmi}{\mathrm{i}}
\newcommand{\rmd}{\mathrm{d}}
\newcommand{\Nf}{N_\mathrm{f}}

\usepackage{xcolor}
\usepackage[normalem]{ulem}

\begin{document}

\title{%
Fractionalized fermionic multicriticality in anisotropic Kitaev spin-orbital liquids%
}

\author{Max Fornoville}

\affiliation{Institut f\"ur Theoretische Physik and W\"urzburg-Dresden Cluster of Excellence ct.qmat, TU Dresden, 01062 Dresden, Germany}

\affiliation{Max Planck Institute for Solid State Research, Heisenbergstra\ss{}e 1, 70569 Stuttgart, Germany}

\affiliation{School of Natural Sciences, Technische Universit\"at M\"unchen, 85748 Garching, Germany}

\author{Lukas Janssen}

\affiliation{Institut f\"ur Theoretische Physik and W\"urzburg-Dresden Cluster of Excellence ct.qmat, TU Dresden, 01062 Dresden, Germany}

\begin{abstract}
We study the low-temperature phase diagram of quantum Kitaev-Heisenberg spin-orbital models with XXZ anisotropy on the honeycomb lattice.
Within a parton mean-field theory, we identify three different quantum phases, distinguished by their symmetries.
Besides a disordered spin-orbital liquid with unbroken $\mathrm{U}(1) \times \mathds Z_2$ spin rotational symmetry, there are two orbital liquid phases characterized by spin long-range order. In these phases, the spin rotational symmetry is spontaneously broken down to residual U(1) and $\mathds Z_2$ symmetries, respectively.
The symmetric spin-orbital liquid features three flavors of linearly dispersing gapless Majorana fermions. 
In the symmetry-broken phases, one of the three Majorana excitations remains gapless, while the other two acquire a band gap.
The transitions from the symmetric to the symmetry-broken phases are continuous and fall into the fractionalized Gross-Neveu-$\mathds Z_2^*$ and Gross-Neveu-SO(2)$^*$ universality classes, respectively.
The transition between the ordered phases is discontinuous.
Using a renormalization group analysis based on the epsilon expansion, we demonstrate that the triple point in the phase diagram features fractionalized fermionic multicriticality with emergent SO(3) symmetry.
\end{abstract}
%

\date{August 19, 2025}

\maketitle

\section{Introduction}
\label{sec:intro}

Spin-orbital liquids are long-range-entangled quantum phases of matter that arise in frustrated magnetic insulators with fluctuating spin and orbital degrees of freedom.
Recently, a number of exactly~\cite{chulliparambil20, chulliparambil21, natori23, keskiner23, akram23, majumder24} or nearly-exactly~\cite{seifert20, jin22, nica23, vijayvargia23, natori24} solvable microscopic models realizing such phases have been proposed.
A key simplification of these models is that the vortices arising from fractionalization of the microscopic degrees of freedom remain static across the phase diagram, leaving the fermionic partons as the only fluctuating excitations. This generalizes the central idea behind the solution of the spin-$1/2$ Kitaev model on the honeycomb lattice~\cite{kitaev06} to systems with additional degrees of freedom.
While no specific material realizations of these phases are currently known, a microscopic mechanism capable of generating the required interactions was recently proposed~\cite{churchill25}.

In contrast to the spin-$1/2$ Kitaev model, the enlarged local Hilbert space resulting from the additional degrees of freedom permits extra terms in the Hamiltonian that preserve the static nature of the vortices.
In Kugel-Khomskii-type Kitaev spin-orbital models~\cite{seifert20, chulliparambil21}, for instance, interactions that couple only spin degrees of freedom and act trivially on the orbital degrees of freedom are permitted.
If sufficiently large, such terms can induce long-range order in the spin sector, while the orbital degrees of freedom remain liquid-like, preserving the long-range entanglement of the ground state.
On the honeycomb lattice, the inclusion of an SU(2) symmetric Heisenberg spin interaction has been shown to induce a transition from the symmetric Kitaev spin-orbital liquid to a symmetry-broken orbital liquid that features N\'eel antiferromagnetic order in the spin sector.
The transition has been argued to be continuous and to represent a realization of a fractionalized fermionic quantum critical point~\cite{seifert20, ray21}.

\begin{figure}[b!]
\includegraphics[width=\linewidth]{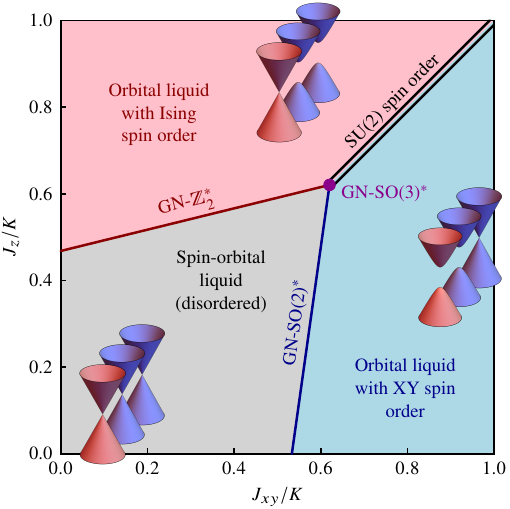}
\caption{%
Phase diagram of Kitaev-Heisenberg-XXZ spin-orbital model in the plane spanned by $J_{xy}$ and $J_z$ from Majorana mean-field theory. Single (double) lines indicate continuous (discontinuous) transitions. The multicritical point GN-SO(3)$^*$ (purple dot) features emergent SO(3) symmetry.
The low-energy spectra of the Majorana fermion flavors in the different phases are sketched in the insets. Blue cones indicate Majorana flavors related by $\mathrm{U}(1)$ spin rotational symmetry, while the red cones indicate the respective remaining third flavor.
}
\label{fig:phasediagram}
\end{figure}

In the present work, we consider a Kitaev-Heisenberg spin-orbital model with XXZ anisotropy in the spin sector using parton mean-field theory and renormalization group  (RG) theory.
Our results are summarized in Fig.~\ref{fig:phasediagram}.
We identify three different quantum phases, distinguished by their symmetries.
Besides a disordered spin-orbital liquid with unbroken $\mathrm{U}(1) \times \mathds Z_2$ spin rotational symmetry, there are two orbital liquid phases characterized by spin long-range order. In these phases, the spin rotational symmetry is spontaneously broken down to residual U(1) and $\mathds Z_2$ symmetries, respectively.
The symmetric spin-orbital liquid features three flavors of linearly dispersing gapless Majorana fermions. 
In the symmetry-broken phases, one of the three Majorana excitations remains gapless, while the other two acquire a band gap.
The symmetry-breaking quantum phase transitions are continuous and can be described in terms of effective relativistic Gross-Neveu-type field theories in $2+1$ space-time dimensions. We analyze the corresponding critical behaviors in terms of an epsilon expansion around the upper critical space-time dimension of four.
In particular, we show that the triple point in the phase diagram corresponds to a fractionalized fermionic multicritical point described by the Gross-Neveu-SO(3) field theory. In the vicinity of the multicritical point, the spin anisotropy becomes irrelevant in the RG sense, leading to multicritical behavior with emergent SO(3) symmetry.
%
%
This result should be contrasted with the multicritical behavior of unfrustrated spin-only models.
In this case, an anisotropy in the spin interaction is relevant in the RG sense and the multicritical point does not feature emergent spin rotational symmetry~\cite{pelissetto02, calabrese03, binder21, rong23, hasenbusch23, hasenbusch24}.

The remainder of this paper is organized as follows: In Sec.~\ref{sec:model}, we introduce the model in both the spin-orbital and Majorana fermion formulations.
The ground states in various perturbative limits are discussed in Sec.~\ref{sec:limits}.
Sec.~\ref{sec:mft} presents our results from the Majorana mean-field theory.
In Sec.~\ref{sec:rg}, we analyze the effects of beyond-mean-field fluctuations using a RG analysis based on the epsilon expansion.
Finally, our conclusions are summarized in Sec.~\ref{sec:conclusions}.

\section{Model}
\label{sec:model}

\subsection{Spin-orbital formulation}

We consider a Kugel-Khomskii-type spin-orbital model on the honeycomb lattice, with Kitaev orbital and XXZ spin exchange interaction, defined by the Hamiltonian
\begin{align} \label{eq:model}
\mathcal H & = - K \sum_{\langle ij \rangle_{\gamma}} \vec{\sigma}_i \cdot \vec{\sigma}_j \otimes \tau_i^\gamma \tau_j^\gamma
+ J_{xy} \sum_{\langle ij \rangle} \sum_{\alpha = x,y} \sigma^\alpha_i \sigma^\alpha_j \otimes \mathds{1}_i \mathds{1}_j
\nonumber \\ &\quad
+ J_z \sum_{\langle ij \rangle} \sigma^z_i \sigma^z_j \otimes \mathds{1}_i \mathds{1}_j
+ \ \dots\,,
\end{align}
with ferromagnetic $(-K) < 0$ and antiferromagnetic $J_{xy}, J_z > 0$.
In the above, $\langle ij \rangle_\gamma$ with $\gamma = x,y,z$ denote the three inequivalent nearest-neighbor bonds on the honeycomb lattice.
The Pauli matrices $\vec \sigma_i = (\sigma^\alpha_i) = (\sigma^x_i, \sigma^y_i, \sigma^z_i)$ act on spin degrees of freedom at site $i$, while the Pauli matrices $(\tau^\gamma_i) = (\tau^x_i, \tau^y_i, \tau^z_i)$ and the identity matrix $\mathds{1}_i$ act on orbital degrees of freedom.
For $J_{xy} \neq J_{z}$, the second and third terms in Eq.~\eqref{eq:model} introduce an XXZ anisotropy in spin space, reducing the SU(2) spin rotational symmetry to a residual $\mathrm{U}(1) \times \mathds Z_2$ symmetry.
The inclusion of XXZ anisotropy is motivated, on one hand, by its relevance to various honeycomb magnets~\cite{das21, maksimov22, bose23, halloran23}.
On the other hand, XXZ anisotropy enables tuning between the easy-axis ($J_z > J_{xy}$) and easy-plane ($J_z < J_{xy}$) regimes, potentially giving rise to multicritical behavior near the higher-symmetry subspace $J_z = J_{xy}$.
In particular, we are interested whether the SU(2) spin rotational symmetry can emerge at low energy as a consequence of strong fluctuations near the multicritical point, despite being absent at the microscopic scale. In the context of the discussion on the multicritical behavior, we will therefore consider also additional terms consistent with the microscopic symmetries of the model. These are represented by the dots in Eq.~\eqref{eq:model} and will be specified later.
In addition to the $\mathrm{U}(1) \times \mathds Z_2$ spin rotational symmetry, the model features a $C_3^*$ symmetry that combines $2\pi/3$ rotations in lattice and orbital space~\cite{janssen19}.

\subsection{Majorana representation}

Important insights into the structure of the model can be obtained by rewriting Eq.~\eqref{eq:model} in terms of a Majorana representation~\cite{yao11,chulliparambil20, seifert20}.
To this end, we introduce six Majorana fermions $(b^\gamma_i)= (b^x_i, b^y_i, b^z_i)$ and $\vec c_i  = (c^\alpha_i) = (c^x_i, c^y_i, c^z_i)$ at each lattice site $i$. The spin-orbital matrices are then represented as
\begin{align}
\sigma_i^\alpha \otimes \mathds{1}_i & \equiv \frac{1}{2} \vec c_i^\top L^\alpha \vec c_i, &
\sigma_i^\alpha \otimes \tau^\gamma_i & \equiv - \rmi b^\gamma_i c^\alpha_i,
\end{align}
where the $3 \times 3$ matrices $(L^\alpha)_{\beta \gamma} = - \rmi \epsilon_{\alpha\beta\gamma}$ are the three generators of SO(3) in the fundamental representation.
The six Majorana fermions act in an extended eight-dimensional local Hilbert space. The physical four-dimensional local Hilbert space consists of those states that are invariant under the $\mathds{Z}_2$ gauge transformation
\begin{align}
D_i = \rmi b^x_i b^y_i b^z_i c^x_i c^y_i c^z_i,
\end{align}
i.e., $D_i$ acts as identity operator on the physical Hilbert space.
Analogous to the spin-$1/2$ case~\cite{kitaev06}, the spin-orbital matrices, as well as any operators constructed from them, including the Hamiltonian, commute with $D_i$, ensuring that the physical Hilbert space is preserved.

In the Majorana representation, the Hamiltonian in Eq.~\eqref{eq:model} reads
\begin{align} \label{eq:model-majorana}
\mathcal H & = K \sum_{\langle i j \rangle_\gamma} u_{ij} \rmi \vec c_i^\top \vec c_j
+ \frac{J_{xy}}{4} \sum_{\langle ij \rangle} \sum_{\alpha = x,y} \vec c_i^\top L^\alpha \vec c_i \vec c_j^\top L^\alpha \vec c_j
\nonumber \\ &\quad
+ \frac{J_{z}}{4}  \sum_{\langle ij \rangle} \vec c_i^\top L^z \vec c_i \vec c_j^\top L^z \vec c_j + \ \dots \,,
\end{align}
where we have defined the $\mathds{Z}_2$ lattice gauge field $u_{ij} = \rmi b_i^\gamma b_j^\gamma$ on the nearest-neighbor links $\langle ij \rangle_\gamma$.
Importantly, the $u_{ij}$'s commute with each other and with the Hamiltonian.
Energy eigenstates are therefore characterized by fixed gauge-field configurations.
The first term in Eq.~\eqref{eq:model-majorana} represents a hopping term for the Majorana fermions in the background of the static gauge field. 
The second and third terms correspond to nearest-neighbor interactions between the Majorana fermions.

\section{Perturbative limits}
\label{sec:limits}

\subsection{$\boldsymbol{J_{xy}, J_z \ll K}$}
\label{subsec:weak-coupling}

For $J_{xy} = J_z = 0$ [and excluding the additional terms represented by the dots in Eq.~\eqref{eq:model-majorana}], the model can be solved exactly. In this case, the ground state is expected to reside in the zero-flux sector~\cite{lieb94, chulliparambil21}, allowing us to assume $u_{ij} = +1$ without loss of generality.
The low-energy spectrum then consists of three flavors of itinerant Majorana fermions $(c^x_i, c^y_i, c^z_i)$ with the band gaps closing at the corners $\mathbf K$ and $\mathbf K'$ of the hexagonal Brillouin zone.
The three flavors are energetically degenerate as a consequence of the enhanced SU(2) spin rotational symmetry for $J_{xy} = J_z = 0$.
The state represents a quantum spin-orbital liquid, and when a topologically nontrivial gap is opened, it realizes the $\nu = 3$ case in Kitaev's 16-fold way of anyon theories~\cite{kitaev06, chulliparambil20}.

For finite $J_{xy}, J_z > 0$, the last two lines in Eq.~\eqref{eq:model-majorana} introduce interactions between the Majorana fermions, which spoils the exact solvability of the model. For small $J_{xy}$ and $J_z$, however, the ground state can be understood within a perturbative analysis.
In particular, vortex excitations remain static and gapped.
Moreover, the four-fermion interactions parametrized by $J_{xy}$ and $J_z$ in Eq.~\eqref{eq:model-majorana} are perturbatively irrelevant in the RG sense.
This implies that the microscopic symmetries remain unbroken in the ground state, and the spin-orbital liquid extends to an entire phase for finite $J_{xy}$ and $J_z$, featuring three flavors of gapless Majorana fermions, provided $J_{xy}$ and $J_z$ stay below a certain finite threshold.
However, in contrast to the SU(2)-symmetric case, only two of the three flavors are energetically degenerate for $J_{xy} \neq J_z$, due to the reduced $\mathrm{U}(1) \times \mathds{Z}_2$ spin rotational symmetry. Nevertheless, all three flavors remain gapless throughout the phase, as mandated by symmetry.

\subsection{$\boldsymbol{J_{xy}, J_z \gg K}$}
\label{subsec:strong-coupling}

The symmetry-broken phases occurring in the strong-coupling regimes $J_{xy}, J_z \gg K$ can already be understood using a simple mean-field decoupling within the spin-orbital formulation of the model. Assuming the mean-field ansatz $\sigma_i^\alpha \simeq \langle \sigma_i^\alpha \rangle = (-1)^i m_\text{stagg}^\alpha$ with antiferromagnetic order parameter $(m_\text{stagg}^\alpha) = (m_\text{stagg}^{xy} \cos \theta, m_\text{stagg}^{xy} \sin \theta , m_\text{stagg}^z)$ in the spin sector yields
\begin{align}
\mathcal H & \simeq K \left[ (m_\text{stagg}^{xy})^{2} + (m_\text{stagg}^z)^2 \right] \sum_{\langle ij \rangle_\gamma} \tau_i^\gamma \tau_j^\gamma 
\nonumber \\ &\quad \label{eq:spin-orbital-mft}
- \frac{3 N}{2} \left[ J_{xy} (m_\text{stagg}^{xy} )^2 + J_z (m_\text{stagg}^z)^2 \right],
\end{align}
where $N$ corresponds to the number of unit cells.

In the strong easy-axis limit $J_{z} \gg J_{xy}, K$, Eq.~\eqref{eq:spin-orbital-mft} suggests Ising antiferromagnetic order in the spin sector, characterized by $ (m_\text{stagg}^\alpha) = (0, 0, \pm 1)$ with spontaneously selected sign, breaking the $\mathds{Z}_2$ part of the spin rotational symmetry.
However, with the background spin long-range order, the first term in Eq.~\eqref{eq:spin-orbital-mft} represents an effective pseudospin-1/2 Kitaev model for the remaining fluctuating orbital degrees of freedom. This implies that two out of the three Majorana fermion flavors acquire a band gap as a consequence of the spin rotational symmetry breaking, while the third one remains gapless and corresponds to the orbital fluctuations.
The ground state for $J_{z} \gg J_{xy}, K$ thus features a $\nu = 1$ orbital liquid accompanied by background Ising antiferromagnetic spin order.

In the strong easy-plane limit $J_{xy} \gg J_{z}, K$, by contrast, Eq.~\eqref{eq:spin-orbital-mft} suggests XY antiferromagnetic order in the spin sector characterized by $(m_\text{stagg}^\alpha) = (\cos \theta, \sin \theta, 0)$ with spontaneously selected polar angle $\theta \in [0, 2\pi)$, breaking the U(1) part of the spin rotational symmetry.
The first term in Eq.~\eqref{eq:spin-orbital-mft} again represents an effective pseudospin-1/2 Kitaev model, such that the ground state for $J_{xy} \gg J_{z}, K$ features a $\nu = 1$ orbital liquid with a single gapless Majorana fermion flavor and XY antiferromagnetic spin order.

\section{Majorana mean-field theory}
\label{sec:mft}

\subsection{Method}

A key simplification of the Majorana interactions parametrized by $J_{xy}$ and $J_z$  in Eq.~\eqref{eq:model-majorana} is the fact that these additional terms still commute with the $u_{ij}$'s, such that the gauge field remains static across the phase diagram.
This property is also essential for realizing the unconventional symmetry-breaking phases, in which the Majorana fermions remain deconfined, cf.~Sec.~\ref{subsec:strong-coupling}.

The model has previously been studied in the SU(2) symmetric case for $J_{xy} = J_z \geq 0$ (excluding the additional terms represented by the dots) using Majorana mean-field theory and density matrix renormalization group (DMRG) calculations~\cite{seifert20}. Additionally, a bilayer variant of the model for $J_{xy} = J_z$ has been studied using sign-problem-free quantum Monte Carlo simulations~\cite{liu22, liu24}.
For $J_{xy} = J_z$, the phase diagram consists of only two phases, namely, the SU(2) symmetric Kitaev spin-orbital liquid for small spin exchange interaction, and a Kitaev orbital liquid with background N\'eel antiferromagnetic spin order for large interaction. The transition is continuous and has been argued to realize an instance of a fractionalized fermionic quantum critical point.
The DMRG calculations~\cite{seifert20} revealed that the ground state remains flux-free across the phase diagram. This implies that we may set $u_{ij} = + 1$ in Eq.~\eqref{eq:model-majorana} without loss of generality, at least for $J_{xy} = J_{z}$. Moreover, the previous results suggest that the natures of the phases and the transition are qualitatively well captured by Majorana mean-field theory.

\begin{figure*}[tb!]
\includegraphics[width=\textwidth]{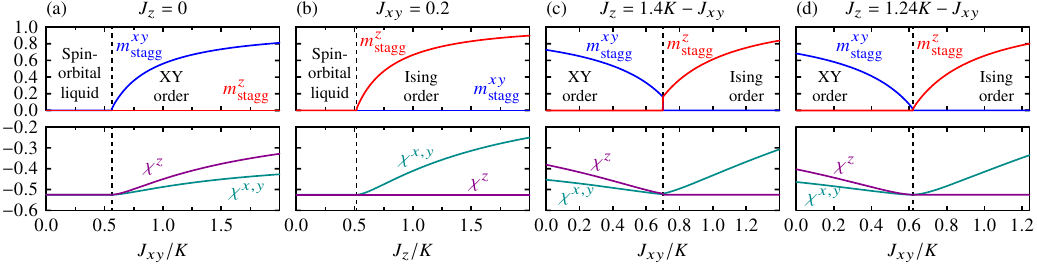}
\caption{%
(a)~Staggered magnetizations $m_\text{stagg}^{xy}$, $m_\text{stagg}^z$ (top) and bond variables $\chi^{x,y}$, $\chi^{z}$ (bottom) as function of $J_{xy}$ for $J_z = 0.2$, from Majorana mean-field theory. Dashed line indicates continuous Gross-Neveu-SO(2)$^*$ transition at $J_{xy}^\text{crit} = 0.561 K$.
(b)~Same as (a), but as function of $J_{z}$ for $J_{xy} = 0.2$ across the continuous Gross-Neveu-$\mathds{Z}_2^*$ transition at $J_z^\text{crit} = 0.517 K$.
(c)~Same as (a), but for $J_{z} = 1.4 K - J_{xy}$ across the discontinuous order-to-order transition at $(J_{xy}^\text{crit},J_z^\text{crit}) = (0.7 K , 0.7K )$.
(d)~Same as (a), but for $J_{z} = 1.24 K - J_{xy}$ across the Gross-Neveu-SO(3)$^*$ multicritical point at $(J_{xy}^\text{crit},J_z^\text{crit}) = (0.620 K , 0.620 K)$.
}
\label{fig:order-parameter}
\end{figure*}

Here, we study the quantum phase diagram for general $J_{xy}$ and $J_z$ within Majorana mean-field theory, with the discussion of the effects of beyond-mean-field fluctuations postponed to Sec.~\ref{sec:rg}.
We consider the on-site parameter
%
$m_i^\alpha = \langle \frac12 \vec c_i^\top L^\alpha \vec c_i \rangle$,
%
which corresponds to the local magnetization at site $i$, and the link variables
%
$\chi_{ij}^\alpha = \rmi \langle c_i^\alpha c_j^\alpha \rangle$
%
on nearest-neighbor bonds $\langle ij \rangle$.
Our mean-field ansatz then reads
\begin{align}
\mathcal H_\text{MF} & = 
\left(K - J_{xy} \chi^z - J_z \chi^y \right) \sum_{\langle ij \rangle} \rmi c^x_i c_j^x 
\nonumber \\ &\quad
+ \left(K - J_{xy} \chi^z - J_z \chi^x \right) \sum_{\langle ij \rangle} \rmi c^y_i c_j^y
\nonumber \\ &\quad
+ \left(K - J_{z} \left(\chi^x  + \chi^y \right) \right) \sum_{\langle ij \rangle} \rmi c_i^z c_j^z
\displaybreak[0] 
\nonumber \\ &\quad
+ J_{xy} \sum_{\langle ij \rangle}\sum_{\alpha = x,y}  \biggl( \frac12 m_i^\alpha \vec c_j^\top L^\alpha \vec c_j + \frac12 \vec c_i^\top L^\alpha \vec c_i m_j 
\nonumber \\ &\quad\qquad
 + \chi^z \chi^\alpha - m_i^\alpha m_j^\alpha\biggr)
\displaybreak[0]
\nonumber \\ &\quad
+ J_{z} \sum_{\langle ij \rangle} \biggl( \frac12 m_i^z \vec c_j^\top L^z \vec c_j + \frac12 \vec c_i^\top L^z \vec c_i m_j^z 
\nonumber \\ &\quad\qquad
+ \chi^x \chi^y - m_i^z m_j^z  \biggr),
\end{align}
where we have assumed that lattice translational and rotational symmetries remain unbroken. This implies that the link variables $\chi_{ij}^\alpha = \chi^\alpha$ are uniform across all nearest-neighbor bonds $\langle ij \rangle$.
Furthermore, it implies that the magnetization $m_i$ is uniform across the unit cells, but possibly different on the two sublattices $A$ and $B$ of the honeycomb lattice, $m_i^\alpha = m_A^\alpha = \text{const.}$ for $i \in A$ and $m_j^\alpha = m_B^\alpha = \text{const.}$ for $j \in B$, respectively.
Moreover, we have assumed that the ground state resides in the flux-free sector, allowing us to set $u_{ij} = +1$ in Eq.~\eqref{eq:model-majorana} without loss of generality, in agreement with the DMRG calculations for $J_x = J_{xy}$~\cite{seifert20}.
This leaves us with nine mean-field parameters $m^\alpha_A$, $m^\alpha_B$, and $\chi^\alpha$, $\alpha = x,y,z$.

To minimize the mean-field energy, we numerically solve the self-consistency equations 
\begin{align}
m^\alpha_A & = \frac{2}{N} \sum_{\mathbf k \in \text{BZ}/2} \langle \vec c_{\mathbf k,A}^{\dagger} L^\alpha \vec c_{\mathbf k,A} \rangle, 
\displaybreak[0] \\
m^\alpha_B & = \frac{2}{N} \sum_{\mathbf k \in \text{BZ}/2} \langle \vec c_{\mathbf k,B}^{\dagger} L^\alpha \vec c_{\mathbf k,B} \rangle,
\displaybreak[0] \\
\chi^\alpha & = \frac{\rmi}{3N} \sum_{\mathbf k \in \text{BZ}/2} \left[f(\mathbf k) \langle \vec c^\dagger_{\mathbf k, A} \vec c_{\mathbf k, A} \rangle - f^\ast(\mathbf k) \langle \vec c^\dagger_{\mathbf k, B} \vec c_{\mathbf k, B} \rangle \right],
\end{align}
where $f(\mathbf k) = 1 + \rme^{- \rmi \mathbf k \cdot \mathbf a_1} + \rme^{-\rmi \mathbf k \cdot \mathbf a_2}$, with $\mathbf a_1$ and $\mathbf a_2$ the primitive vectors on the honeycomb lattice, $N$ corresponds to the number of real-space unit cells, and the momentum summation is over half of the first Brillouin zone (BZ/2).
Initial values for the mean-field parameters are randomly chosen and then iterated until convergence is reached.
As a cross-check, the energy of the resulting state is compared with the energy of the symmetric solution $m_A^\alpha = m_B^\alpha = 0$ in order to ensure that the ground state is reached.

\subsection{Phase diagram}

The top row of Fig.~\ref{fig:order-parameter} shows the evolution of the mean-field order parameters $m^{xy}_\text{stagg} = \frac12 \sqrt{(m^x_A - m^x_B)^2 + (m^y_A - m^y_B)^2}$ and $m^z_\text{stagg} = |m^z_A - m^z_B|/2$ as functions of $J_{xy}/K$ and $J_{z}/K$. The bottom row shows the corresponding bond variables $\chi^\alpha$, $\alpha = x,y,z$.
For small $J_{xy}$ and $J_z$ [Figs.~\ref{fig:order-parameter}(a,b)], the system feature a fully symmetric ground state characterized by $m^{xy}_\text{stagg} = m^z_\text{stagg} = 0$ and $\chi^x = \chi^y = \chi^z$.

Upon increasing $J_{xy}$ for fixed small $J_z$ [Fig.~\ref{fig:order-parameter}(a)], a staggered magnetization in the XY plane $m^{xy}_\text{stagg}$ develops. The new ground state for large $J_{xy}$ spontaneously breaks the U(1) spin symmetry, in agreement with the strong-coupling analysis discussed in Sec.~\ref{subsec:strong-coupling}.
The quantum phase transition between the symmetric spin-orbital liquid and the orbital liquid with background XY spin order is continuous. The presence of gapless fermions at the transition, however, makes the quantum critical point nontrivial, already at the mean-field level. In Fig.~\ref{fig:order-parameter}(a), this is evident from the finite slope of the order parameter at the transition, indicating that the quantum critical point does not belong to the usual XY universality class. In the XY case, the order parameter would be expected to exhibit a square-root dependence, $m_\text{stagg}^{xy} \propto \sqrt{J_{xy} - J_{xy}^\text{crit}}$, at the mean-field level.
Below, we argue that the transition between the symmetric spin-orbital liquid and the orbital liquid with background XY spin order falls into the fractionalized Gross-Neveu-SO(2)$^*$ universality class.%
\footnote{Following established conventions~\cite{isakov12, whitsitt16, schuler16, seifert20, ray21}, we mark fractionalized universality classes with an asterisk ($*$). The fractionalized transitions differ from their nonfractionalized counterparts by the absence of gauge-dependent operators in the spectrum, while featuring additional states associated with topological degeneracies~\cite{seifert20}.}

\begin{figure*}[tb!]
\includegraphics[width=\textwidth]{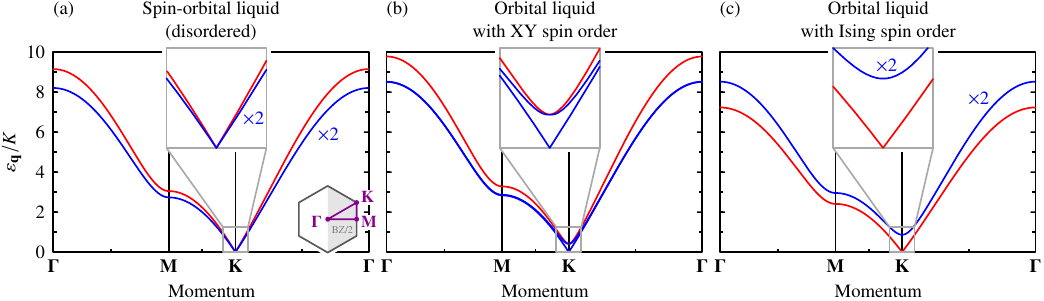}
\caption{%
(a)~Spectrum of itinerant Majorana fermions for $(J_{xy},J_z) = (0.5,0.2)K$ in the disordered spin-orbital liquid phase along a high-symmetry path in the half of the first Brillouin zone (BZ/2), indicated in the lower right inset. For $J_{xy} > J_z$, the two degenerate bands (blue) associated with the U(1) spin rotational symmetry have lower energy than the nondegenerate band (red). All three bands become gapless at the $\mathbf K$ point, see upper inset.
(b)~Same as (a), but for $(J_{xy},J_z) = (0.6,0.2)K$ in the orbital liquid phase with XY antiferromagnetic spin order. The two formerly-degenerate bands (blue) now split at the $\mathbf K$ point, with one of the bands remaining gapless, while the other one acquiring a band gap, together with the nondegenerate band (red).
(c)~Same as (a), but for $(J_{xy},J_z) = (0.2,0.6)K$ in the orbital liquid phase with Ising antiferromagnetic spin order. The two degenerate bands (blue) acquire a band gap at the $\mathbf K$ point, while the nondegenerate band (red) remains gapless.
}
\label{fig:spectra}
\end{figure*}

Upon increasing $J_z$ for fixed small $J_{xy}$ [Fig.~\ref{fig:order-parameter}(b)], a staggered magnetization along the Ising axis $m^{z}_\text{stagg}$ develops. The new ground state for large $J_{z}$ spontaneously breaks the $\mathds Z_2$ spin symmetry, in agreement with the strong-coupling analysis discussed in Sec.~\ref{subsec:strong-coupling}.
Again, the corresponding quantum phase transition is continuous and nontrivial. Below, we argue that it falls into the fractionalized Gross-Neveu-$\mathds Z_2^*$ universality class.

Upon increasing $J_{xy}$ for fixed large $J_z$ [Fig.~\ref{fig:order-parameter}(c)], a direct transition from the orbital liquid with XY spin order to the orbital liquid with Ising spin order is realized. For the present model without additional symmetry-allowed terms in the Hamiltonian (denoted by the dots in Eq.~\eqref{eq:model}), the transition occurs along the high-symmetry line $J_{xy} = J_z$. If further symmetry-allowed terms are taken into account, the transition line as function of $J_{xy}$ and $J_{z}$ for fixed values of the additional parameters will shift away from the high-symmetry subspace. However, if the additional parameters are small, the transition should be expected to remain direct and discontinuous.

The magnitude of the order parameter jump at the first-order transition varies with the distance to the triple point, where three symmetry-distinct phases meet. In the absence of additional symmetry-allowed terms in the Hamiltonian, we locate the triple point at $J_{xy} = J_z = 0.620K$. When tuning $J_{xy}$ and $J_z$ directly through this point, the order-to-order transition becomes continuous, as shown in Fig.~\ref{fig:order-parameter}(d). Below, we argue that this transition belongs to the Gross-Neveu-SO(3)$^*$ universality class, consistent with the earlier analysis along the $J_{xy} = J_z$ line, where the $\mathrm{U}(1) \times \mathds{Z}_2$ spin rotational symmetry enhances to SU(2)~\cite{seifert20, ray21}.

\subsection{Majorana excitation spectrum}

Figure~\ref{fig:spectra} shows the positive part of the fermion excitation spectrum $\varepsilon_{\mathbf q}$ from Majorana mean-field theory for representative values of $J_{xy}$ and $J_z$ in the three different phases. The corresponding negative part of the spectrum is determined by particle-hole symmetry.
The symmetric spin-orbital liquid phase [Fig.~\ref{fig:spectra}(a)] is characterized by two degenerate bands (blue), associated with the U(1) spin rotational symmetry, and a nondegenerate band (red). For $J_{xy} > J_z$, the two degenerates band have lower energy than the nondegenerate band across the Brillouin zone, except at its corner $\mathbf K$. At this point, both bands become gapless.
In the orbital liquid with XY spin order [Fig.~\ref{fig:spectra}(a)], the U(1) spin rotational symmetry is spontaneously broken, leading to a splitting of the two formely-degenerate bands (blue). Near the $\mathbf{K}$ point, one of the bands opens a finite gap in the spectrum along with the nondegenerate band (red), while the remaining band continues to be gapless.
In the orbital liquid with Ising spin order [Fig.~\ref{fig:spectra}(a)], the U(1) spin rotational symmetry is unbroken and the two associated bands (blue) stay degenerate, but open a finite gap in the spectrum near the $\mathbf K$ point.
By contrast, the nondegenerate band (red) in this phase has lower energy than the two degenerate bands and remains gapless at the $\mathbf K$ point.

\section{Renormalization group analysis}
\label{sec:rg}

\subsection{Low-energy field theory}

The presence of gapless Majorana fermions in the spectrum invalidates the conventional Landau paradigm of phase transitions, which considers only the fluctuations of the order parameters.
To assess the natures of the different transitions, we derive a low-energy field theory in the spirit of a gradient expansion~\cite{herbut06, seifert20}.
To this end, we rewrite the lattice Majorana fermion $c_{i,s}^\alpha(\tau)$, where $i$ denotes the unit cell, $s = A, B$ the sublattice, and $\alpha = x,y,z$ the flavor, in terms of complex continuum fermion fields $\psi_{s}^\alpha(\tau,\mathbf x)$ at imaginary time $\tau$ and position $\mathbf x$ as
\begin{align}
	c_{i,s}^\alpha(\tau) = \psi^\alpha_s(\tau, \mathbf x) \rme^{\rmi \mathbf K \cdot \mathbf x} + \text{H.c.}\,,
\end{align}
and expand for small momenta $\mathbf q = \mathbf k - \mathbf K$ near the single nodal point $\mathbf K$.
By symmetry, the resulting low-energy action must take the form
\begin{align}
S_{\psi} & = \int \! \rmd^2 \mathbf x\, \rmd \tau\, \Biggl\{ \bar \psi (\gamma^\mu \otimes \mathds{1}_3) \partial_\mu \psi
\nonumber \\ &\quad \label{eq:action-fermionic}
+ \tilde J_{xy} \sum_{\alpha = x,y} [\bar\psi (\mathds 1_2 \otimes L^\alpha) \psi]^2
+ \tilde J_{z} [\bar\psi (\mathds 1_2 \otimes L^z) \psi]^2
 \Biggr\}\,,
\end{align}
with four-fermion couplings $\tilde J_{xy} \propto a J_x/K$ and $\tilde J_z \propto a J_z/K$, where $a$ is the lattice constant.
Here, the gamma matrices $\gamma^\mu$ satisfy the Euclidean Clifford algebra $\{\gamma^\mu, \gamma^\nu\} = 2 \delta^{\mu\nu} \mathds{1}_2$.
In the above and in what follows, the summation convention over repeated space-time indices $\mu = 0,1,2$ is assumed.
The $2 \times 2$ gamma matrices act on the sublattice degree of freedom, while the $3 \times 3$ matrices $L^x$, $L^y$, and $L^z$ [generators of SO(3)] act on the flavor degree of freedom of the six-component spinors $\psi = (\psi_A^x, \psi_A^y, \psi_A^z, \psi_B^x, \psi_B^y, \psi_B^z)$ and $\bar\psi = \psi^\dagger \gamma^0$.
In Eq.~\eqref{eq:action-fermionic}, the fermion fields $\psi^{\alpha}_s$ have been rescaled so that the prefactor of the kinetic term is unity for all three flavors $\alpha = x,y,z$.

To analyze the effective field theory defined by Eq.~\eqref{eq:action-fermionic}, it is convenient to introduce the XY order-parameter field $\vec\phi = (\phi^x, \phi^y)$ and the Ising order-parameter field $\phi^z$ via a Hubbard-Stratonovich transformation, leading to the fermion-boson action
\begin{align}
S_{\psi\phi} & = \int \! \rmd^2 \mathbf x\, \rmd \tau \Biggl\{ \bar \psi (\gamma^\mu \otimes \mathds{1}_3) \partial_\mu \psi
\nonumber \\ &\quad
+ g_{xy} \sum_{\alpha = x,y} \phi^\alpha \bar\psi (\mathds{1}_2 \otimes L^\alpha) \psi 
+ g_{z} \phi^z \bar\psi (\mathds{1}_2 \otimes L^z) \psi 
\nonumber \\ &\quad
+ \frac{r_{xy}}{2} \sum_{\alpha = x,y} (\phi^{\alpha})^2 
+ \frac{r_z}{2} (\phi^{z})^2
 \Biggr\}\,,
\label{eq:action-fermion-boson}
\end{align}
where the Yukawa couplings $g_{xy}$, $g_z$ and the tuning parameters $r_{xy}$, $r_z$ are related to the four-fermion couplings as $g^2_{xy}/(2r_{xy}) = \tilde J_{xy}$ and $g_z^2/(2 r_z) = \tilde J_{z}$.
At low energy, additional terms not present in the microscopic action may be generated from fluctuations. In particular, kinetic terms for the order parameters and bosonic self-interactions are allowed by symmetry,
\begin{align}
S_\phi & = \int \! \rmd^2 \mathbf x\, \rmd \tau \Biggl\{
- \frac{1}{2} \sum_{\alpha = x,y} \phi^{\alpha} \partial_\mu^2 \phi^\alpha
- \frac{1}{2} \phi^{z} \partial_\mu^2 \phi^z
\nonumber \\ &\quad
+ \lambda_{xy} \left[\sum_{\alpha = x,y}(\phi^\alpha)^2\right]^2
+ \lambda_z (\phi^z)^4
\nonumber \\ &\quad
+ 2 \lambda_{xyz} \sum_{\alpha = x,y}(\phi^\alpha)^2 (\phi^z)^2
\Biggr\}\,.
\end{align}
Analogous to the fermion fields, the order-parameter fields have been rescaled such that the kinetic-term prefactor is unity for both $\vec \phi = (\phi^x, \phi^y)$ and $\phi^z$.
Higher-order terms are irrelevant in the RG sense, at least within the epsilon expansion discussed below.
The full low-energy effective action is thus given by 
\begin{align} \label{eq:action-full}
S = S_{\psi\phi} + S_\phi.
\end{align}
In the low-energy effective theory, the $\mathrm{U}(1)$ spin rotational symmetry of the microscopic model is realized as
\begin{align}
\mathrm{U}(1)&:&
\psi & \mapsto \rme^{\rmi \vartheta (\mathds{1}_2 \otimes L^z)} \psi, &
\bar\psi & \mapsto \bar \psi \rme^{ - \rmi \vartheta (\mathds{1}_2 \otimes L^z)} \psi, \nonumber \\
&&
\vec \phi & \mapsto R \vec \phi, &
\phi^z & \mapsto \phi^z,
\end{align}
with arbitrary angle $\vartheta \in [0, 2\pi)$ and SO(2) rotation matrix $R(\vartheta) = \rme^{-\rmi \vartheta \sigma_y}$. The XY order parameter $\vec \phi = (\phi^x, \phi^y)$ transforms as an SO(2) vector under this symmetry transformation.
The $\mathds{Z}_2$ spin symmetry is realized as
\begin{align}
\mathds{Z}_2&:&
\psi & \mapsto \rme^{\rmi \pi (\mathds{1}_2 \otimes L^x)} \psi, &
\bar\psi & \mapsto \bar \psi \rme^{ - \rmi \pi (\mathds{1}_2 \otimes L^x)} , \nonumber \\
&&
\vec \phi & \mapsto (\phi^x, -\phi^y), &
\phi^z & \mapsto - \phi^z.
\end{align}
Enhancing the $\mathrm{U}(1) \times \mathds{Z}_2$ spin rotational symmetry to SU(2) requires that three parameters in the effective low-energy theory vanish, namely $\Delta g = g_{xy} - g_z$, $\Delta \lambda_{xy} = \lambda_{xy} - \lambda_{xyz}$, and $\Delta \lambda_{z} = \lambda_{z} - \lambda_{xyz}$.

We emphasize that the low-energy model defined in Eq.~\eqref{eq:action-full} incorporates all symmetry-allowed interactions that are potentially relevant in the RG sense, including those arising from the terms represented by the dots in Eq.~\eqref{eq:model}, as long as the latter are small compared to the flux gap.
The number of parameters that need to vanish in order to reach the subspace with enhanced SU(2) symmetry is therefore higher in the effective low-energy model than in the microscopic model (without additional terms).
However, we demonstrate below that all three parameters $\Delta g$, $\Delta \lambda_{xy}$, and $\Delta \lambda_z$ naturally flow to zero in the long-wavelength limit, provided that $r_{xy}$ and $r_z$ are tuned to their respective critical values. This leads to an emergent SU(2) spin rotational symmetry at the multicritical point, even when the additional $\mathrm{U}(1) \times \mathds{Z}_2$-symmetric terms, represented by the dots in Eq.~\eqref{eq:model}, are finite and not explicitly fine-tuned.

\subsection{Flow equations}

Generalizing the low-energy field theory defined in Eq.~\eqref{eq:action-full} to $D$ space-time dimensions, the five independent couplings in the theory have the tree-level scaling dimensions
\begin{align}
[g_{xy}] & = [g_z] = \frac{4-D}{2}, \\
[\lambda_{xy}] & = [\lambda_z] = [\lambda_{xyz}] = 4-D
\end{align}
The theory thus features a unique upper critical space-time dimension of four, allowing us to compute the universal critical behavior within a standard epsilon expansion, with $\epsilon = 4-D$.

Integrating out the high-energy fluctuations with $d+1$ momentum $q$ in the shell between $\Lambda/b$ and $\Lambda$ leads to the flow equations
\begin{align} \label{eq:flow-r-1}
\frac{\rmd r_{xy}}{\rmd \ln b} & = (2 - \eta_{\phi}^{xy}) r_{xy} - \frac{4 \Nf}{3} g_{xy}^2 + \frac{16 \lambda_{xy}}{1+ r_{xy}} + \frac{4 \lambda_{xyz}}{1+r_z}, \\
\frac{\rmd r_{z}}{\rmd \ln b} & = (2 - \eta_{\phi}^{z}) r_{z} - \frac{4 \Nf}{3} g_{z}^2 + \frac{12 \lambda_{z}}{1+ r_{z}} + \frac{8 \lambda_{xyz}}{1+r_{xy}},
\end{align}
for the two control parameters $r_{xy}$ and $r_z$, where $\Nf = 3$ corresponds to the number of two-component Dirac fermion flavors,
\begin{align}
\frac{\rmd g_{xy}^2}{\rmd \ln b} & = (\epsilon - \eta_\phi^{xy} - \eta_\psi^{xy} - \eta_\psi^z) g_{xy}^2 - \frac{2 g_{xy}^4}{1 + r_{xy}}, \\
\frac{\rmd g_{z}^2}{\rmd \ln b} & = (\epsilon - \eta_\phi^{z} - 2\eta_\psi^{xy}) g_{z}^2 - \frac{2 g_{z}^4}{1 + r_{z}},
\end{align}
for the two Yukawa couplings, and
\begin{align}
\frac{\rmd \lambda_{xy}}{\rmd \ln b} & = (\epsilon - 2\eta_\phi^{xy}) \lambda_{xy} + \frac{\Nf}{3} g_{xy}^4 - \frac{40 \lambda_{xy}^2}{(1+r_{xy})^2} - \frac{4 \lambda_{xyz}^2}{(1+r_z)^2}, \\
\frac{\rmd \lambda_{z}}{\rmd \ln b} & = (\epsilon - 2\eta_\phi^{z}) \lambda_{z} + \frac{\Nf}{3} g_{z}^4 - \frac{36 \lambda_{xy}^2}{(1+r_{z})^2} - \frac{8 \lambda_{xyz}^2}{(1+r_{xy})^2}, \\
\frac{\rmd \lambda_{xyz}}{\rmd \ln b} & = (\epsilon - \eta_\phi^{xy} - \eta_z) \lambda_{xyz} + \frac{\Nf}{3} g_{xy}^2 g_z^2 - \frac{16 \lambda_{xy} \lambda_{xyz}}{(1+r_{xy})^2}
\nonumber \\ & \quad
- \frac{12 \lambda_z \lambda_{xyz}}{(1+r_z)^2} - \frac{16 \lambda_{xyz}^2}{(1+r_{xy})(1+r_z)},
\end{align}
for the three boson couplings, with the boson anomalous dimensions
\begin{align} \label{eq:eta-phi}
\eta_\phi^{xy} & = \frac{2\Nf}{3} g_{xy}^2, &
\eta_\phi^{z} & = \frac{2\Nf}{3} g_{z}^2
\end{align}
of the XY and Ising order-parameter fields $\vec \phi = (\phi^{x}, \phi^y)$ and $\phi^z$, respectively, and the fermion anomalous dimensions
\begin{align} \label{eq:eta-psi}
\eta_\psi^{xy} & = \frac{g_{xy}^2}{2(1+r_{xy})^2} + \frac{g_z^2}{2(1+r_z)^2}, &
\eta_\psi^z & = \frac{g_{xy}^2}{(1+r_{xy})^2}
\end{align}
of the fermion fields $(\psi^x_s, \psi^y_s)$ and $\psi^z_s$, respectively.
Note that the fermion two-point correlation functions are not invariant under $\mathbb Z_2$ gauge transformations, and therefore the fermion anomalous dimensions are not physically observable in our microscopic realization.
In the above flow equations and anomalous dimensions, we have rescaled the tuning parameters as $(r_{xy}, r_z)/\Lambda^2 \mapsto (r_{xy}, r_z)$ and the couplings as $(g^2_{xy}, g^2_z)/(8 \pi^2 \Lambda^\epsilon) \mapsto (g^2_{xy}, g^2_z)$ and $(\lambda_{xy}, \lambda_z, \lambda_{xyz})/(8 \pi^2 \Lambda^\epsilon) \mapsto (\lambda_{xy}, \lambda_z, \lambda_{xyz})$.
Note that the two distinct fermion anomalous dimensions $\eta_\psi^{xy}$ and $\eta_\psi^z$, as well as the two distinct boson anomalous dimensions $\eta_\phi^{xy}$ and $\eta_\phi^z$ arise due to the reduced $\mathrm{U}(1) \times \mathds{Z}_2$ spin rotational symmetry.
In the respective limits, Eqs.~\eqref{eq:flow-r-1}--\eqref{eq:eta-psi} simplify to known results from the literature:
For $r_{xy} = r_z$, $g_{xy} = g_z$, and $\lambda_{xy} = \lambda_z = \lambda_{xyz}$, we recover the flow equations of the Gross-Neveu-SO(3) model~\cite{seifert20, ray21}.
For $g_{xy} = g_z = 0$, the equations reduce to those of the purely bosonic model with $\mathrm{U}(1) \times \mathds{Z}_2$ symmetry~\cite{pelissetto02, calabrese03, herbutbook}.

\subsection{Quantum critical behavior}

\begin{figure*}[t]
\includegraphics[width=\textwidth]{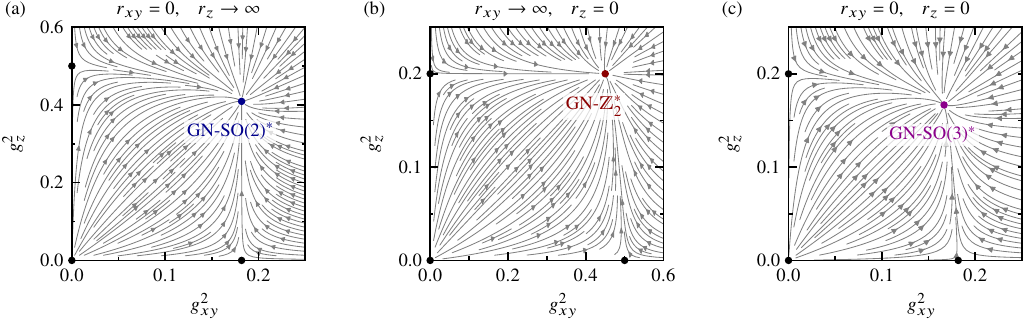}
\caption{%
(a)~Renormalization group flow in the plane spanned by the Yukawa couplings $g^2_{xy}$ and $g_z^2$ for $r_{xy} = 0$ and $r_z \to \infty$, from one-loop epsilon expansion, showing the infrared stable fixed point GN-SO(2)$^*$, describing the Gross-Neveu-SO(2)$^*$ transition.
(b)~Same as (a), but for $r_{xy} \to \infty$ and $r_z = 0$, showing the infrared stable fixed point GN-$\mathds{Z}_2^*$, describing the Gross-Neveu-$\mathds{Z}_2^*$ transition.
(c)~Same as (a), but for $r_{xy} = 0$ and $r_z = 0$, showing the infrared stable fixed point GN-SO(3)$^*$, describing the Gross-Neveu-SO(3)$^*$ multicritical point.
}
\label{fig:rgflow}
\end{figure*}

\subsubsection{Spin-orbital liquid to XY antiferromagnet}

As $J_{xy}$ increases for a fixed small $J_z$, the mean-field analysis in Sec.~\ref{sec:mft} suggests a direct continuous transition from the disordered Kitaev spin-orbital liquid to an orbital liquid exhibiting XY antiferromagnetic spin order, see Fig.~\ref{fig:order-parameter}(a).
At the transition, the XY order-parameter field $\vec \phi = (\phi^x, \phi^y)$ becomes gapless, while the Ising order-parameter field $\phi^z$ remains fully gapped. As a result, the low-energy effective field theory for this transition follows from the full effective action in Eq.~\eqref{eq:action-full} in the limit $r_{xy} = 0$ and $r_z \to \infty$, yielding the Gross-Neveu-SO(2) model.
The corresponding RG flow diagram in the plane spanned by $g_{xy}^2$ and $g_z^2$ is shown in Fig.~\ref{fig:rgflow}(a). The flow exhibits a unique infrared stable fixed point, with a single RG relevant direction along the $r_{xy}$ axis.
Consequently, the continuous spin-orbital-liquid-to-XY-order transition belongs to the universality class defined by this fixed point.
It is located at
\begin{align}
(g_{xy}^{\ast 2}, g_{z}^{\ast 2} ) = (\tfrac{2}{11}, \tfrac{17}{114}) \epsilon + \mathcal O(\epsilon^2)
\end{align}
and
\begin{align}
(\lambda_{xy}^\ast, \lambda_z^\ast, \lambda_{xyz}^\ast) = (\tfrac{3 + \sqrt{649}}{880}, \tfrac{4212\sqrt{649}-85860}{177408}, \tfrac{3 \sqrt{649} - 39}{352} ) \epsilon + \mathcal O(\epsilon^2),
\end{align}
where we have set $\Nf = 3$, relevant for the lattice model defined in Eq.~\eqref{eq:model}.
Following the terminology of Refs.~\cite{isakov12, whitsitt16, schuler16, seifert20}, we refer to the corresponding fractionalized universality class as Gross-Neveu-SO(2)$^*$ and label the fixed point GN-SO(2)$^*$ in Fig.~\ref{fig:rgflow}(a).
The transition is characterized by the order-parameter anomalous dimension
\begin{align}
\eta_{\phi}^{xy} = \frac{4}{11} \epsilon + \mathcal O(\epsilon^2)
\end{align}
and the correlation-length critical exponent
\begin{align}
1/\nu  = 2 - \frac{23 + \sqrt{649}}{55}\epsilon + \mathcal O(\epsilon^2),
\end{align}
which are both clearly distinct from the corresponding values in the classical XY universality class~\cite{herbutbook, chester20}.
The remaining exponents can be obtained via the usual hyperscaling relations. For the scaling of the order-parameter as a function of $J_{xy} - J_{xy}^\text{crit}$, for instance, we obtain
\begin{align}
1/\beta = 2-\frac{\sqrt{649}-12}{55}  \epsilon + \mathcal O(\epsilon^2)
\end{align}
for $\Nf = 3$.
In the large-$\Nf$ limit, we obtain $1/\beta =  2-\epsilon + \mathcal O(\epsilon^2, 1/\Nf)$, consistent with the mean-field value $\beta = 1$ in $D= 3$ space-time dimensions. This explains the finite slope of the order parameter at the critical point in Fig.~\ref{fig:order-parameter}(a).

Two remarks are in order:
First, in our calculation, the Gross-Neveu-SO(2)$^*$ fixed point is located at finite $g_z^{*2}$, even though the Ising field $\phi^z$ is gapped at the transition and could, in principle, be integrated out from the start. We stress, however, that $g_z^{*2}$ in this case affects only nonuniversal quantities, such as the fixed-point location, and does not influence any universal property. In particular, we have explicitly verified that integrating out the gapped $\phi^z$ field from the outset leaves all critical exponents unchanged.
Second, we note that the critical exponents of the Gross-Neveu-SO(2)$^*$ universality class differ from those of the conventional Gross-Neveu-XY class~\cite{zerf17}, even though both involve spontaneous $\mathrm{SO}(2) \simeq \mathrm{U}(1)$ symmetry breaking and---for appropriately chosen fermion numbers $\Nf$---share the same field content. This difference is fundamental: it originates from qualitative changes in the fermionic spectrum, which lead to distinct low-energy field theories and, consequently, different universality classes. In the Gross-Neveu-XY case, all fermion modes are gapped in the symmetry-broken phase, giving a semimetal-insulator transition. In our model, the fermionic spectrum remains gapless across the transition. This is due to the spin-1 matrices $L^\alpha$, $\alpha = x,y$, present in the Yukawa interaction term parametrized by $g_{xy}$ in Eq.~\eqref{eq:action-fermion-boson}, each of which features a zero eigenvalue. As a consequence, one out of three bands in the fermionic spectrum remains gapless in the XY antiferromagnetic phase, cf.\ Fig.~\ref{fig:spectra}(b), in contrast to the situation in the conventional Gross-Neveu-type models~\cite{janssen14, zerf17}.

\subsubsection{Spin-orbital liquid to Ising antiferromagnet}

As $J_z$ increases for a fixed small $J_{xy}$, the mean-field analysis suggests a direct continuous transition from the disordered Kitaev spin-orbital liquid to an orbital liquid exhibiting $\mathds{Z}_2$ antiferromagnetic spin order, characterized by the Ising order-parameter field $\phi^z$. At the transition, the XY order-parameter field $\vec \phi = (\phi^x, \phi^y)$ remains fully gapped, such that the corresponding low-energy effective field theory follows from Eq.~\eqref{eq:action-full} in the limit $r_{xy} \to \infty$ and $r_z = 0$.
The corresponding RG flow diagram in the plane spanned by $g_{xy}^2$ and $g_z^2$ is shown in Fig.~\ref{fig:rgflow}(b). Analogously to the XY transition discussed above, the flow exhibits a unique infrared stable fixed point, with a single RG relevant direction that is now aligned along the $r_z$ axis. It defines the Gross-Neveu-$\mathds Z_2^*$ universality class and is located at
\begin{align}
(g_{xy}^{\ast 2}, g_{z}^{\ast 2} ) = (\tfrac{9}{20}, \tfrac{1}{5}) \epsilon + \mathcal O(\epsilon^2)
\end{align}
and
\begin{align}
(\lambda_{xy}^\ast, \lambda_z^\ast, \lambda_{xyz}^\ast) & = (\tfrac{25920 \sqrt{145} - 299295}{72000}, \tfrac{1 + \sqrt{145}}{360}, \tfrac{3 \sqrt{145} - 30}{50}) \epsilon
\nonumber \\ & \quad
+ \mathcal O(\epsilon^2),
\end{align}
where we have again set $\Nf = 3$, relevant for the lattice model in Eq.~\eqref{eq:model}.
The corresponding critical exponents read
\begin{align}
\eta_{\phi}^{z} = \frac{2}{5} \epsilon + \mathcal O(\epsilon^2)
\end{align}
and
\begin{align}
1/\nu  = 2 - \frac{\sqrt{145} + 13}{30}\epsilon + \mathcal O(\epsilon^2),
\end{align}
which are again different from the values in the corresponding classical Ising universality class~\cite{herbutbook, chang25}. 
We emphasize that the fractionalized Gross-Neveu-$\mathds Z_2^*$ universality class defined by the above fixed point is also distinct from the recently much-studied conventional Gross-Neveu-Ising universality class~\cite{janssen14, zerf17, ihrig18, huffman20, tabatabaei22, erramilli23}. 
In analogy to the transition between the spin-orbital liquid to the XY antiferromagnet discussed above, the difference is due to the spin-1 matrix $L^z$ present in the Yukawa interaction term parametrized by $g_z$ in Eq.~\eqref{eq:action-fermion-boson}, which features a zero eigenvalue. As a consequence, one out of the three bands in the fermion spectrum remains gapless in the Ising antiferromagnetic phase, cf.\ Fig.~\ref{fig:spectra}(c), in contrast to the situation in the conventional Gross-Neveu-Ising model.
At the transition, this gapless mode is fully decoupled from the remaining degrees of freedom, implying that the corresponding Fermi velocity is not renormalized, in agreement with our mean-field results, cf.~Fig.~\ref{fig:spectra}(c).

\subsubsection{Multicritical point}

We now discuss the quantum multicritical behavior in the vicinity of the triple point at which the disordered spin-orbital liquid and the XY and Ising antiferromagnetic phases meet. Without the additional terms represented by the dots in Eq.~\eqref{eq:model}, the triple point is located within the higher-symmetric subspace featuring SU(2) spin rotational symmetry at $(J_{xy}^\text{crit}, J_z^\text{crit}) = (0.620 K, 0.620 K)$, within the mean-field calculation. By continuity, analogous triple points should be expected to exist also in the presence of additional terms compatible with the $\mathrm U(1) \times \mathds Z_2$ spin rotational symmetry of the microscopic model, such as XXZ spin exchange interactions on next-nearest neighbor bonds $\langle\!\langle ij \rangle\!\rangle$,
\begin{align} \label{eq:next-nearest-neighbor}
\mathcal H_{2} = \sum_{\langle\!\langle ij \rangle\!\rangle} \left( J_{2}^{xy} 
\sum_{\alpha = x,y} \sigma^\alpha_i \sigma^\alpha_j \otimes \mathds{1}_i \mathds{1}_j
+ J_{2}^{z} \sigma^z_i \sigma^z_j \otimes \mathds{1}_i \mathds{1}_j \right),
\end{align}
at least as long as $J_2^{xy}$ and $J_{2}^{z}$ are small compared to $J_{xy}$ and $J_z$.
For finite $J_2^{xy} \neq J_2^z$, the triple points will generically no longer be located within the SU(2) invariant subspace.
We show below, however, that all SU(2) breaking terms compatible with the $\mathrm U(1) \times \mathds Z_2$ microscopic symmetry of the model are irrelevant at the multicritical fixed point. This implies that SU(2) spin rotational symmetry emerges at low energy at any multicritical point between the disordered spin-orbital liquid and the XY and Ising antiferromagnetic phases, even if the symmetry is explicitly broken at the microscopic level in the presence of the additional terms represented by the dots in Eq.~\eqref{eq:model}, such as those given by the next-nearest-neighbor terms with finite $J_2^{xy} \neq J_2^{z}$ in Eq.~\eqref{eq:next-nearest-neighbor}.
To demonstrate this, consider the effective action in Eq.~\eqref{eq:action-full} in the limit in which both order parameters are critical, $r_{xy} = r_z = 0$.
The corresponding RG flow diagram in the plane spanned by $g_{xy}^2$ and $g_z^2$ is shown in Fig.~\ref{fig:rgflow}(c). The flow exhibits a unique infrared stable fixed point, with two RG relevant directions along $(1, 1)$ and $(1,-2)$ in the $r_{xy}$-$r_z$ plane. It is located at
\begin{align}
(g_{xy}^{\ast 2}, g_{z}^{\ast 2} ) = (\tfrac{1}{6}, \tfrac{1}{6}) \epsilon + \mathcal O(\epsilon^2)
\end{align}
and
\begin{align}
(\lambda_{xy}^\ast, \lambda_z^\ast, \lambda_{xyz}^\ast) = (\tfrac{3+9 \sqrt{5}}{792}, \tfrac{3+9 \sqrt{5}}{792}, \tfrac{3+9 \sqrt{5}}{792}) \epsilon + \mathcal O(\epsilon^2),
\end{align}
where we have again set $\Nf = 3$, relevant for the lattice model in Eq.~\eqref{eq:model}. 
Importantly, the fixed point governing the multicritical behavior is located in the SU(2) invariant subspace given by $\Delta g = g_{xy} - g_z =0$, $\Delta \lambda_{xy} = \lambda_{xy} - \lambda_{xyz} = 0$, and $\Delta \lambda_{z} = \lambda_{z} - \lambda_{xyz} = 0$.
This implies that any small perturbation that breaks the SU(2) spin rotational symmetry to $\mathrm U(1) \times \mathds Z_2$, such as those given by the next-nearest-neighbor terms with finite $J_2^{xy} \neq J_2^{z}$ in Eq.~\eqref{eq:next-nearest-neighbor}, is irrelevant in the RG sense, as advertised above. As a consequence of the emergent SU(2) symmetry, the multicritical point falls into the universality class of the Gross-Neveu-SO(3)$^*$ model, with anomalous dimensions~\cite{seifert20, ray21}
\begin{align}
\eta_\phi^{xy} = \eta_{\phi}^{z} =  \frac{\epsilon}{3} + \mathcal O(\epsilon^2).
\end{align}
The correlation-length scaling depends on the direction at which the triple point is approached. 
The smallest positive eigenvalue of the stability matrix at the multicritical fixed point is along the $(1,1)$ direction in the $r_{xy}$-$r_z$ plane, which corresponds to the direction within the SU(2) invariant subspace.
As a consequence, for a generic direction within the $\mathrm{U}(1) \times \mathds Z_2$ invariant parameter space, the correlation-length exponent at the multicritical point is the same as that of the SU(2) invariant Gross-Neveu-SO(3)$^*$ model~\cite{seifert20, ray21},
\begin{align}
1/\nu  = 2 - \frac{9 + 5 \sqrt{5}}{22} \epsilon + \mathcal O(\epsilon^2).
\end{align}
If, by contrast, the multicritical point is approached along the specific direction corresponding to $(1,-2)$ in the $r_{xy}$-$r_z$ plane, the correlation length diverges with correlation-length exponent given by
\begin{align}
1/\nu \Big|_{(1,-2)}  = 2-\frac{4+\sqrt{5}}{11} \epsilon + \mathcal O(\epsilon^2).
\end{align}

\section{Conclusions}
\label{sec:conclusions}

In this work, we have studied the phases and phase transitions of Kitaev-Heisenberg spin-orbital models with XXZ anisotropy in the spin sector.
The models fall into the larger class of Kugel-Khomskii-type spin-orbital models featuring liquid-like ground states with long-range entanglement~\cite{chulliparambil20, chulliparambil21, natori23, keskiner23, akram23, majumder24, seifert20, jin22, nica23, vijayvargia23, natori24}.
In addition to the Kitaev spin-orbital interaction, we have taken anisotropic XXZ spin-only interactions into account. Importantly, the spin-only interactions preserve the static nature of the vortices in the liquid-like ground state. As a consequence, the long-range orders induced in the spin sector preserve the liquid nature of the orbital degrees of freedom with long-range quantum entanglement.
We identify two distinct orbital-liquid phases with antiferromagnetic spin long-range order at sizable spin-only interactions.
The transitions from the disordered spin-orbital liquid to the antiferromagnetically-ordered orbital liquid phases are continuous and define new fractionalized fermionic universality classes Gross-Neveu-SO(2)$^*$ and Gross-Neveu-$\mathds Z_2^*$, respectively.
At the triple point at which the three phases meet, SU(2) spin rotational symmetry is emergent at low energy. This implies that the multicritical point falls into the higher-symmetric Gross-Neveu-SO(3)$^*$ universality class, which has recently been characterized within a variety of analytical~\cite{seifert20,ray21} and numerical~\cite{liu22,liu24} tools.

Our result that SU(2) emerges at low energy at the multicritical point should be contrasted with the behavior in conventional spin-only models.
In this case, an anisotropy in the spin interaction is relevant in the RG sense and the universal behavior at the triple point is different from those of the SU(2) symmetric Heisenberg universality class~\cite{pelissetto02, calabrese03, binder21, rong23, hasenbusch23, hasenbusch24}.
The emergent symmetry at the multicritical point arises from the fractionalization of the spin degrees of freedom into static gauge fluxes and itinerant fermions. In the presence of the latter, several examples of enhanced symmetries at multicritical points are known in the context of Gross-Neveu-type theories~\cite{roy11,janssen18,roy18}.

A material realization of any of the phases discussed in this work will inevitably involve additional interactions that do not commute with the gauge field, thereby breaking the static nature of the vortices. For instance, in the recently proposed microscopic mechanism for realizing the Kitaev spin-orbital interaction~\cite{churchill25}, a conventional Kugel-Khomskii interaction is also expected to arise. For the future, it would therefore be interesting to study the interplay between the terms that commute with the gauge field, such as the Kitaev spin-orbital and XXZ spin-only interactions discussed in this work, with additional terms that do not commute with the gauge field. The latter will generically favor conventional long-range order, potentially leading to rich phase diagrams including long-range-entangled phases with and without background spin order as well as short-range-entangled phases with long-range spin and orbital orders.

\begin{acknowledgments}

We are grateful to 
Sreejith Chulliparambil,
Urban Seifert,
Hong-Hao Tu,
and 
Matthias Vojta
for insightful discussions and collaborations on related topics.
This work has been supported by the Deutsche Forschungsgemeinschaft (DFG) through SFB 1143 (A07, Project No.\ 247310070), the W\"urzburg-Dresden Cluster of Excellence \textit{ct.qmat} (EXC 2147, Project No.\ 390858490), and the Emmy Noether program (JA2306/4-1, Project No.\ 411750675). M.F. was partly funded by the European Union (ERC-2023-StG, Project No.\ 101115758 -- QuantEmerge). Views and opinions expressed are, however, those of the authors only and do not necessarily reflect those of the European Union or the European Research
Council Executive Agency. Neither the European Union nor the granting authority can be held responsible for them.
\end{acknowledgments}

\paragraph*{Data availability.} The data that support the findings of this article are openly available~\cite{data-availability}.



\bibliographystyle{longapsrev4-2}
\bibliography{frac-mcrit}

\end{document}